# Instability of charge density wave in Kagome magnet FeGe

Ziyuan Chen[1], Xueliang Wu[2], Ruotong Yin[1], Jiakang Zhang[1], Shiyuan Wang[1], Yuanji Li[1], Mingzhe Li[1], Aifeng Wang[2], Yilin Wang[1], Ya-Jun Yan[1*], Dong-Lai Feng[1,3*]

[1] School of Emerging Technology and Department of Physics, University of Science and Technology of China, Hefei, 230026, China

[2] Low temperature Physics Laboratory, College of Physics and Center of Quantum Materials and Devices, Chongqing University, Chongqing 401331, China

[3] National Synchrotron Radiation Laboratory and School of Nuclear Science and Technology, New Cornerstone Science Laboratory, University of Science and Technology of China, Hefei, 230026, China

*Corresponding authors：yanyj87@ustc.edu.cn; dlfeng@ustc.edu.cn

## Abstract

**Kagome metals show rich competing quantum phases due to geometry frustration, flat bands, many-body effects, and non-trivial topology. Recently, a novel charge density wave (CDW) was discovered deep inside the antiferromagnetic phase of FeGe, attracting intense attention due to close relation with magnetism. Here, via a scanning tunneling microscope (STM), we find the 2 × 2 CDW in FeGe is very fragile and can be readily disrupted into the initial 1 × 1 phase; Small √3 × √3 CDW puddles are found to coexist with the 2 × 2 CDW in as-grown samples, and can also be induced in the intermediate process of CDW disruption, which will eventually transform into the initial 1 × 1 phase. Moreover, an exotic intermediate CDW state and standalone CDW nuclei appear unexpectedly during the disruption process. Our first-principle calculations find equal softening of a flat optical phonon mode in a large momentum region around the CDW wave vector, corresponding to numerous competing CDWs with close energies. This might lead to strong instability of the CDW ground state, responsible for STM observations. Our findings provide more novel experimental aspects to understand the CDW in FeGe and suggest FeGe-like Kagome metals are ideal platforms for studying the physics of competing CDW instabilities.**

## I. INTRODUCTION

The concept of charge density wave (CDW) permeates much of condensed matter physics and chemistry. CDW is static modulations of electron density distribution that usually accompanied by periodic lattice distortions [1-3]. The study of its physical properties and underlying mechanism is crucial for further understanding the intrinsic interactions in the system, such as electron-phonon coupling, electron correlation, etc [1-19]. After extensive investigations, several types of CDWs are suggested based on the formation mechanism. Type I is the weak-coupling CDW originating from Fermi surface nesting. It usually occurs in quasi-1D systems, where phonons are only affected (softened) at the CDW wave vector $Q$ that spans the nested portions of Fermi surface near the CDW transition [1,2]. Type II CDW is driven by strong electron-phonon coupling, which was first envisaged by McMillan in 1977, to explain the anomalously large $\Delta/(k_B T_{CDW})$ in 2H-TaSe$_2$ and alike [4]. Here, phonon frequencies soften over a large portion of reciprocal space around the CDW wave vector. Near phase transition, numerous phonon modes attempt to soften and condense simultaneously, resulting in unignorable phononic entropy, which leads to large $\Delta/(k_B T_{CDW})$ and pseudogap-like behavior above $T_{CDW}$ [4-8]. Below $T_{CDW}$, a long-ranged CDW with a fixed wave vector is stabilized. Besides, there are some CDWs that cannot be classified into the above two types, such as those in cuprates [9-14], iron-based superconductors [15-17] and excitonic insulators [18,19]. For the former two systems, strong electron correlation effect is considered to play a dominated role [9-17]; while for the last system, the softened collective modes are excitons or plasmons rather than phonons [18,19].

Kagome metals show rich competing quantum states due to the unique geometry frustration, flat bands, many-body effects, and non-trivial topology, such as quantum spin liquid, exotic CDW orders and unconventional superconductivity [20-23]. It has become an ideal platform for exploring exotic states with strong quantum fluctuations. For example, the CDWs in Kagome metal $AV_3Sb_5$ (A = K, Rb, Cs) have attracted extensive attention because of diverse CDW patterns [24-31] and complex broken symmetries [28-31]. Extensive experimental and theoretical studies on the band structures, phonon spectra, and lattice distortions suggest a probable origin of Fermi surface nesting of van Hove singularities. Recently, a novel CDW order was discovered in a Kagome antiferromagnet FeGe [32-42], with $T_{CDW}$ ~ 110 K, far below the antiferromagnetic (AFM) transition temperature of 410 K [37-40]. The occurrence of a CDW deep inside the AFM phase is very rare, and it also significantly enhances the spin-polarization [37], suggesting close relation between CDW and magnetism. To further understand the CDW mechanism, in this manuscript, we use a scanning tunneling microscope (STM) to study the properties of the CDW state in FeGe. We find the 2 × 2 CDW in FeGe is very fragile and can be readily disrupted into the initial 1 × 1 phase; Small √3 × √3 CDW puddles are found to coexist with the 2 × 2 CDW in as-grown samples, and can also be induced in the intermediate process of CDW disruption, which will eventually transform into the initial 1 × 1 phase. Moreover, an exotic intermediate CDW state and standalone CDW nuclei appear unexpectedly during the disruption process. Our first-principle calculations find softening of a flat optical phonon mode by nearly the same magnitude in a large portion of Brillouin zone (BZ) around the CDW wave vector, which corresponds to numerous competing CDWs with very close energies. This might lead to strong instability of the CDW ground state in FeGe, responsible for the easy CDW disruption and the appearance of novel CDW states. Our results provide more novel experimental aspects to understand the CDW in FeGe.

## II. METHODS
### A. Crystal growth of B35-type FeGe

Single crystals of B35-type FeGe were synthesized by a chemical vapor transport method using $I_2$ as the transport agent. Iron powders (99.99%) and germanium powders (99.999%) were mixed according to the stoichiometric ratio, finely ground in an agate mortar, and then sealed in an evacuated quartz tube with 20 cm in length. The quartz tube was subsequently annealed in a two-zone horizontal furnace with a temperature gradient from 600 °C (source) to 550 °C (sink) for 12 days. FeGe single crystals with a typical size of 1.5 × 1.5 × 3 mm$^3$ can be obtained in the middle of the quartz tube. Subsequently, the as-grown FeGe crystals were post-annealed at 320 °C for 48 h, and then quenched in water. As-grown FeGe crystals were characterized by x-ray diffraction (XRD) measurements using a Rigaku D/max-A x-ray diffractometer with Cu Kα radiation in the range of 10° – 100° with steps of 0.01° at room temperature. Temperature dependent magnetic susceptibilities of as-grown FeGe crystals and those annealed at 320 °C were measured using Quantum Design Magnetic Property Measurement System.

## B. STM measurements

For STM experiments, FeGe crystals were cleaved at 78 K in ultrahigh vacuum with a base pressure better than $1 \times 10^{-10}$ mbar and immediately transferred into a UNISOKU cryogenic STM at $T = 4.5$ K. The STM bias is applied to the sample, with the STM tip grounded. Pt-Ir tips were used after being treated on a clean Au (111) substrate. The d$I$/d$V$ spectra were collected by a standard lock-in technique with a modulation frequency of 731 Hz and a modulation amplitude $\Delta V$ of about 2-30 mV for different conditions.

## C. DFT calculations

The DFT plus Hubbard $U$ calculations are performed using Vienna ab initio simulation package (VASP) [43]. The exchange-correlation potential is treated within the generalized gradient approximation (GGA) of the Perdew-Burke-Ernzerhof variety [44]. The simplified approach introduced by Dudarev et al. (LDAUTYPE=2) is used [45]. We used the experimental lattice parameters of FeGe [36,37]. Phonon calculations are performed in the A-type AFM phase with a $2 \times 2 \times 1$ supercell (with respect to the AFM cell), using both the density-functional-perturbation theory (DFPT) [46] and frozen phonon approaches, combined with the Phonopy package [47]. The two approaches yield identical results. The internal atomic positions of charge dimerized $2 \times 2 \times 2$ and $\sqrt{3} \times \sqrt{3} \times 2$ CDW phases are relaxed with the initial atomic distortions corresponding to the soft optical phonon mode at **L** and **H** points [40], respectively, until the force is less than 0.001 eV/Å for each atom. Integration for the Brillouin zone is done using a Γ-centered $8 \times 8 \times 10$ $k$-point grids and the cutoff energy for plane-wave-basis is set to be 500 eV.

## III. EXPERIMENTAL RESULTS

### A. Crystal structure and characterization of FeGe single crystals

Figure 1(a) shows the XRD measurements of $\theta$-$2\theta$ scan of the *ab* and *bc* planes of FeGe single crystals, sharp (00*l*) or (*l*00) peaks are observed, respectively. The calculated *a*-axis and *c*-axis lattice parameters are 4.98 Å and 4.05 Å, consistent with previous reports of B35-type FeGe [37]. Figure 1(b) shows the room-temperature crystal structure of B35-type FeGe, consisting of alternating stacked $Fe_3Ge$ Kagome layer and $Ge_2$ honeycomb layer. Temperature dependent magnetic susceptibilities of as-grown FeGe crystals and those annealed at 320 °C are displayed in Fig. 1(c), with the CDW transitions indicated by the magenta arrows. As have been reported in our previous article [48], the CDW transition in FeGe is very sensitive to the post annealing process — the weak drop at ~ 100 K for as-grown samples suggests a short-ranged CDW, while the sharper drop at ~ 110 K for post-annealed samples corresponds to the formation of a long-ranged CDW. Benefiting to the post-annealed samples with a long-ranged CDW, we have resolved the CDW superstructure of FeGe [48], which is dominated by a large dimerization of 1/4 of the Ge-1 sites in the Kagome layers along the *c*-axis, as shown in Fig. 2(d). If interested, please see ref. [48] of our previous article to find more details about the annealing treatment regulated CDW and underlying mechanism.

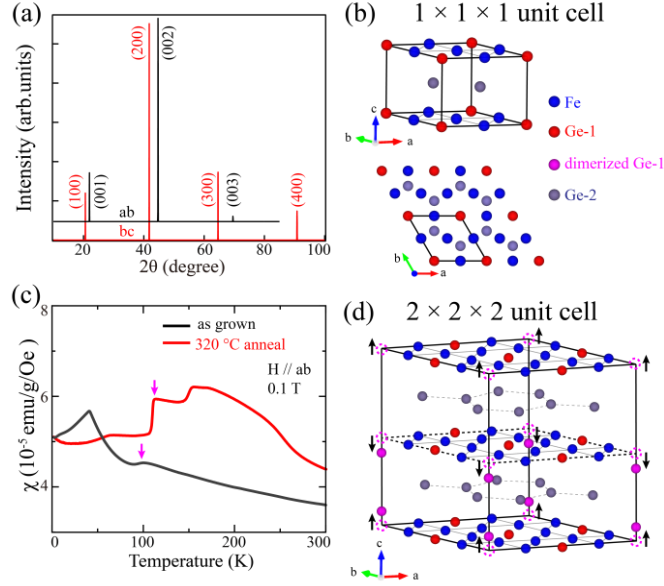

FIG. 1. Crystal structure and characterization of B35-type FeGe single crystals. (a) XRD patterns for the *ab* and *bc* planes of as-grown FeGe crystals. (b) Room-temperature crystal structure of FeGe (before CDW transition). Fe, Ge-1, dimerized Ge-1 and Ge-2 atoms are indicated by the blue, red, magenta and gray spheres, respectively. (c) Temperature dependent in-plane magnetic susceptibilities of as-grown FeGe and FeGe that annealed at 320 °C. The CDW transitions are indicated by the magenta arrows. (d) Crystal structure of FeGe after CDW transition, which is reproduced from ref. [48].

### B. Basic properties of Fe$_3$Ge and Ge$_2$ terminated surfaces of as-grown FeGe crystals

After cleaving FeGe crystals, there should be two kinds of exposed surfaces, the Fe$_3$Ge Kagome layer and the Ge$_2$ honeycomb layer. These two terminated surfaces show distinct topographic images and d$I$/d$V$ spectra as previously reported in ref. [48]. Figures 2(a) and 2(b) show the typical topographic images of Fe$_3$Ge and Ge$_2$ surfaces of as-grown FeGe crystals, there are some native atomic defects on the Fe$_3$Ge surface, while many residual bright clusters distribute randomly on the Ge$_2$ surface, as indicated by the yellow arrows. In the nominated Fe$_3$Ge surface, we have observed occasionally a hexagonal atomic lattice with a special tip condition (inset of Fig. 2(a)), the interatomic spacing is ~ 5 Å, the same as the Ge-1 atomic lattice, further proving the determination of Fe$_3$Ge surface.

From Figs. 2(a), 2(b) and 2(c) that shows a typical d$I$/d$V$ map of the same sample region as in Fig. 2(a) at $V_b$ = 0.3 V, it is found that both the Fe$_3$Ge and Ge$_2$ surfaces have two types of domains, the ideal 1 × 1 Kagome lattice and the 2 × 2 charge modulation, as demonstrated more clearly in Figs. 2(e) and 2(f). The fast-Fourier transformation (FFT) image of Fig. 2(c) is shown in Fig. 2(d), where a set of much broadened spots ($q_{2a}$) locates at half of the wave vectors of underlying lattice ($q_a$), confirming a short-ranged 2 × 2 CDW. We follow the method descried in ref. [49] to calculate the in-plane translational correlation length of the CDW shown in Fig. 2(c) and get a result of correlation length ξ of ~ 1.95 nm. By averaging more sample regions as discussed in Fig. S1 of Supplementary Materials (SM) [50], a value of ξ ~ 2.4 nm is obtained, which is consistent to the estimated values in previous neutron scattering experiment (~ 3.3 nm) [37]. When scrutinizing Figs. 2(e) and 2(f) with an overlaid atomic structure, you can find that in our STM experiments, it is difficult to distinguish different atoms clearly, whether in topography or in d$I$/d$V$ map; and most of the time, what we see is the behavior of the whole Kagome unit cell. This allows us to observe the CDW modulations, but hard to determine the distribution of charge disproportionation between different atoms.

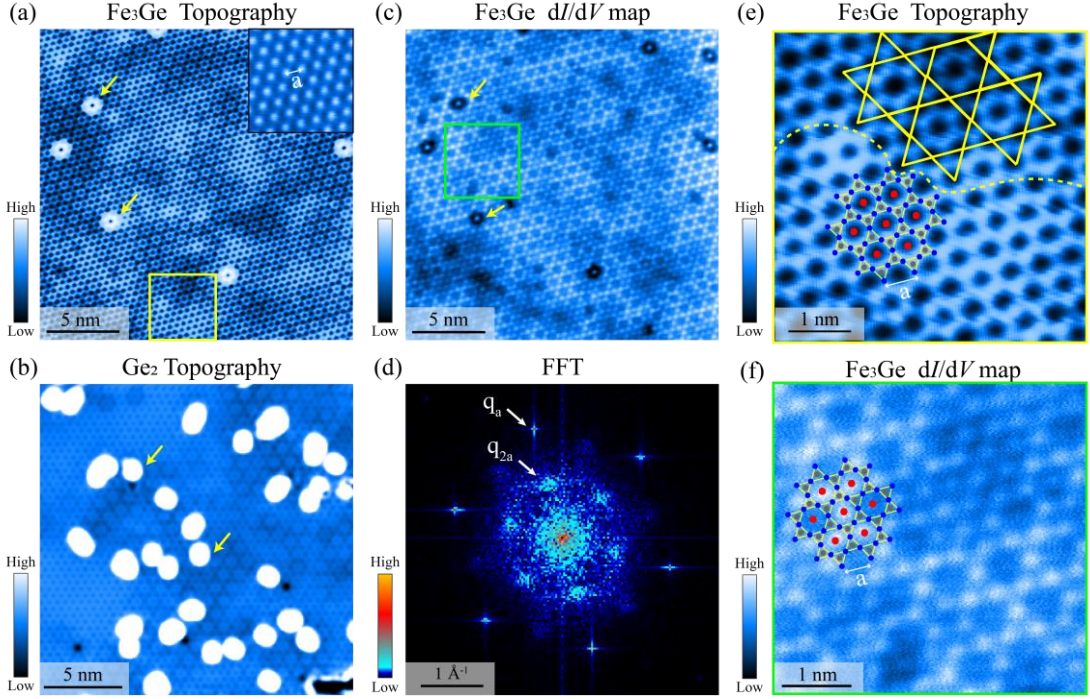

FIG. 2. Short-ranged 2 × 2 CDW in as-grown FeGe. (a),(b) Typical topographic images of the $Fe_3Ge$ and $Ge_2$ terminated surfaces. Inset of panel (a) shows a hexagonal atomic lattice with the interatomic spacing of ∼ 5 Å, consistent with Ge-1 atomic lattice. (c) Typical d$I$/d$V$ map at $V_b$ = 0.3 V, measured in the same sample region shown in panel (a). The yellow arrows in panels (a-c) indicate the native atomic defects in $Fe_3Ge$ layer and the residual clusters in $Ge_2$ layer after cleavage. (d) FFT image of panel (c), with the Bragg spots of underlying lattice and CDW labeled as $q_a$ and $q_{2a}$, respectively. (e),(f) Enlarged images of the areas enclosed by the yellow and green boxes in panels (a) and (c), respectively. The 1 × 1 Kagome lattice is marked out by the small yellow hexagrams with Fe (blue), Ge-1 (red) and Ge-2 (gray) atoms indicated, while the CDW superlattice is indicated by the thick yellow lines in panel (e). The yellow dashed curve in panel (e) shows approximately the boundary between the 1 × 1 and 2 × 2 patterns. Measurement conditions: (a) $V_b$ = 0.3 V, $I_t$ = 300 pA; (b) $V_b$ = 0.2 V, $I_t$ = 100 pA; (c) $V_b$ = 0.3 V, $I_t$ = 300 pA, $\Delta V$ = 30 mV.

### C. Disruption of the short-ranged CDW in as-grown FeGe crystals

More remarkably, we find the CDW in FeGe can be readily disrupted by gentle STM scans. An example is shown in Fig. 3. Figure 3(a) is the initial topographic image taken at a normal condition of $V_b$ = 50 mV and $I_t$ = 20 pA, exhibiting randomly distributed CDW domains as enclosed by the orange dashed curves. This region is then scanned once at a higher bias of $V_b$ = 1 V, and subsequently Fig. 3(b) is obtained at $V_b$ = 0.2 V. It is peculiar that partial CDW domains become smaller or disappear, while the others remain unchanged. By repeating the above process again, the remaining CDW domains are erased gradually and almost disappear after 3-5 cycles, as displayed in Figs. 3(c) and 3(d). Figure S2 of SM in ref. [50] displays the CDW disruption process of another sample region with multiple step terraces. Moreover, we find that without moving the STM tip but just hovering it atop the sample surface with a mildly higher bias can also destroy the CDW in a sizeable nearby region and induce the transformation into the $1 \times 1$ phase. Notably, the CDW can only be erased in the scanned regions or a sizeable nearby region around the hovered STM tip, leaving the untreated region, even a few nanometers away, unaffected.

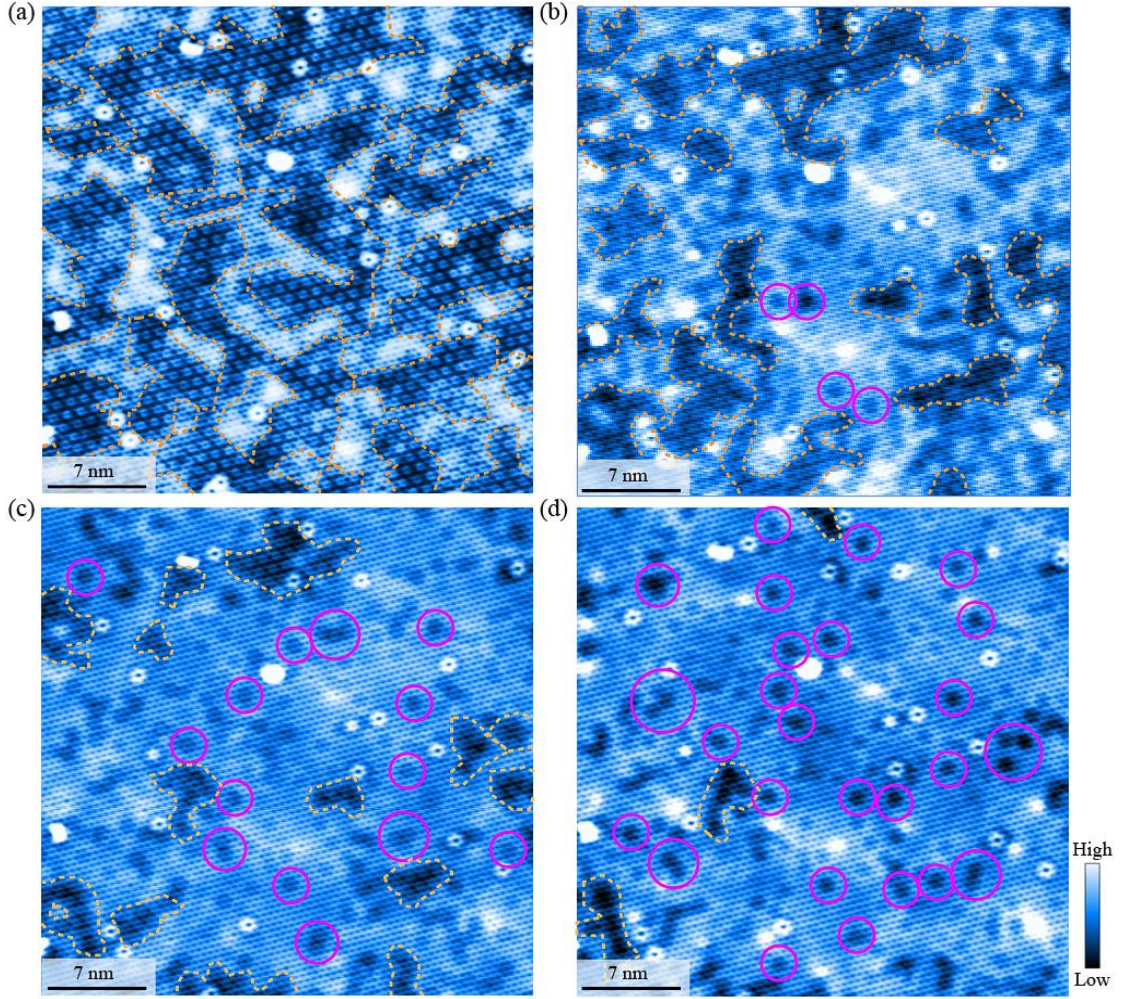

FIG. 3. Stepwise disruption of the short-ranged CDW in FeGe. (a)-(d) Topographic images for progressive disruption stages in the same sample region, acquired after treating the sample region with increasing STM scan cycles at a high bias of $V_b = 1$ V and $I_t = 20$ pA. The CDW domains are enclosed by the orange dashed curves, and the magenta circles in panels (b)-(d) highlight the appearance of individual CDW nucleus. Measurement conditions: (a) $V_b = 50$ mV, $I_t = 20$ pA; (b)-(d) $V_b = 0.2$ V, $I_t = 100$ pA.

### D. Disruption of the long-ranged CDW in annealed FeGe crystals

The above disruption method is not limited to the short-ranged CDW observed in as-grown FeGe, but also works for the long-ranged CDW in annealed FeGe crystals [48,51,52]. With the same method and condition (scanning or hovering STM tip atop at ~ 1 V), we can also disrupt the long-ranged CDW in annealed FeGe crystals. Figure 4 shows an example. Before disruption, the CDW is well-ordered in the whole field of view (Fig. 4(a)), proving the long-ranged nature; but it is broken into several different phases after the disruption process described above (Figs. 4(b) and 4(c)). The CDW remains unaffected in some areas, but becomes weaker or even disappears in the other areas. An intriguing intermediate state appears in some areas as enclosed by the white dashed curves in Figs. 4(b) and 4(c), which still exhibits the same $2 \times 2$ period but the amplitude of charge modulation is significantly reduced compared with the initial CDW. It will eventually transform to the $1 \times 1$ phase. These phenomena are illustrated more clearly in Fig. 4(d), which displays the line profiles of charge distribution along the same spatial trajectory in Figs. 4(a)-4(c). A well-ordered $2 \times 2$ charge modulation with large amplitude is observed along the whole trajectory before CDW disruption (black curve in Fig. 4(d)), but is only kept in part of the trajectory after CDW disruption (left parts of the red and blue curves in Fig. 4(d)). For the red curve, the initial CDW damps gradually across the phase boundary of about 3-4 super-unit-cell in size and stabilizes to the intermediate $2 \times 2$ CDW state with a fixed weak amplitude. For the blue curve, the situation is similar, except that the initial CDW eventually damps to the $1 \times 1$ phase.

As for the structure of the intermediate CDW state, now it is hard to pin down experimentally. Firstly, as shown in Figs. 2(e) and 2(f), our STM study cannot resolve the precise atomic positions so as to judge subtle changes in lattice distortions; Secondly, the region of intermediate CDW is too small to allow the application of other techniques. However, we have a reasonable hypothesis for the origin of intermediate CDW state. The intermediate CDW appears to be a crossover state connecting the long-ranged CDW and the initial $1 \times 1$ phase. As suggested in refs. [48,51,52] and illustrated in Fig. 1, the biggest structural difference between the long-ranged CDW and the initial $1 \times 1$ phase is Ge-1 dimerization. Considering a different magnitude in Ge-1 dimerization, which might be induced locally by internal defects or external interferences, CDW states with different amplitudes, such as the long-ranged CDW, intermediate CDW and CDW in the as-grown FeGe, might be induced. We notice that a similar metastable state has been mentioned in a theoretical article [53]. Anyway, this hypothesis needs further verifications.

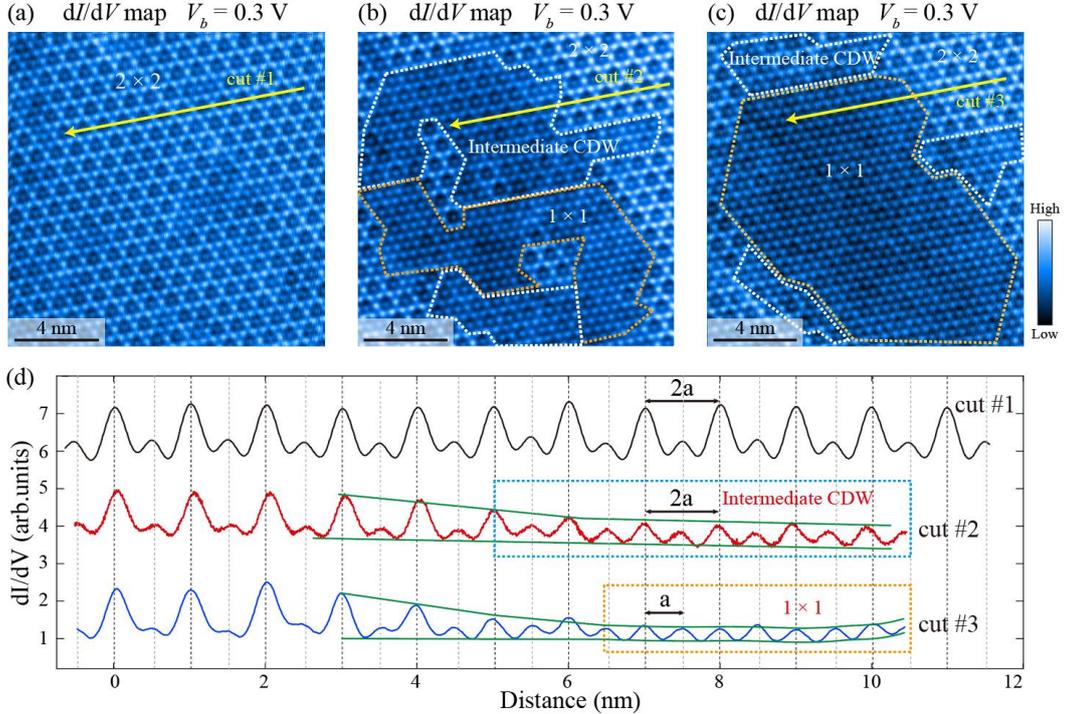

FIG. 4. Stepwise disruption of a long-ranged CDW and the appearance of an intermediate CDW state. (a)-(c) $dI/dV$ maps for the progressive disruption stages in the same sample region, acquired after treating the sample region by hovering STM tip at $V_b = 1$ V and $I_t = 20$ pA. The areas with an intermediate CDW state and without CDW are enclosed by the white and orange dashed curves, respectively. (d) Line profiles collected along cuts #1-#3 as indicated in panels (a)-(c). Measurement conditions: (a)-(c) $V_b$ = 0.3 V, $I_t$ = 300 pA, $\Delta V$ = 30 mV.

### E. More datasets for CDW disruption and possible mechanism

As we know, phase transitions induced by the STM tip are usually considered to be related to several mechanisms, including mechanical deformations, local Joule heating, electric field effect of the tip, and charge injection. To understand the mechanism of CDW disruption in FeGe, we have tried different set point conditions, as shown in Fig. 5. When $|V_b| \geq 0.9$ V, a $I_t$ value of ~ 10 pA is enough to disrupt the CDW, precluding local Joule heating as the main driving force. When $|V_b|$ < 0.5 V, even if we apply a $I_t$ value of 3000 pA (the highest value used in our STM experiments), there is no obvious effect on CDW distribution. This suggests that purely mechanical deformations of the sample induced by pushing the STM tip closer to the sample surface with lower $|V_b|$ and higher $I_t$ are precluded. Considering the existence of a threshold $|V_b|$ for CDW disruption, we believe that the phase transition is driven in a large part by the electric field effect at the apex of the STM tip, since an enough electric field is needed to overcome the energy barrier from the CDW state to the pristine phase. The intense electric field in STM experiments is localized in a sample region with a lateral size of ~ 10 nm under the STM tip, which increases and extends laterally with increasing $|V_b|$. This naturally explains our observations in the left panel of Fig. 5(b) that the region with CDW disruption increases with increasing $|V_b|$. Moreover, we also notice that when $|V_b|$ exceeds the threshold, increasing $I_t$ can accelerate the disruption process quickly. Since increasing $I_t$ can only increase the electric field of the tip slightly, thus, more factors, such as charge injection, may also assist the CDW disruption by inelastic processes. More theoretical and experimental studies are needed to explore the detailed process of CDW disruption. Besides, in our study, we find that the CDW disruption is irreversible, once destroyed, the original 2 × 2 or other CDWs can't be restored by STM scans or tip pulses as usually used in other CDW systems [54], which is also an open question.

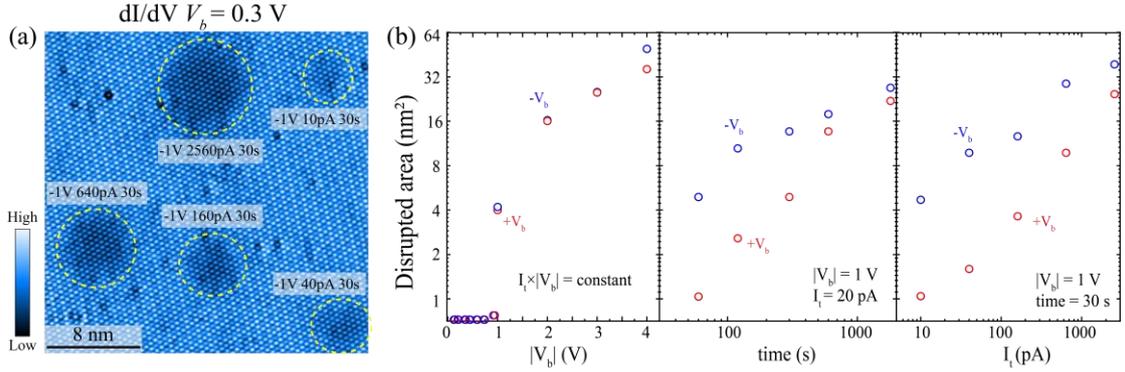

FIG. 5. More datasets for CDW disruption. (a) A typical d$I$/d$V$ map of Fe$_3$Ge layer, illustrating the CDW disruption areas by hovering the STM tip under different set point conditions. The disrupted CDW areas are enclosed by the yellow dashed curves. (b) Statistical graphs of disrupted areas under different set point conditions. Measurement conditions: (a) $V_b$ = 0.3 V, $I_t$ = 300 pA, $\Delta V$ = 30 mV.

### F. Observations of √3 × √3 CDW puddles and CDW nuclei

During STM measurements, we find occasionally small puddles of √3 × √3 charge modulation in as-grown FeGe, either before or during CDW disruption process. An example is shown in Figs.

6(a) and 6(b). The √3 × √3 CDW puddle, as enclosed by the orange dashed curves, coexists with the 2 × 2 CDW (enclosed by the white dashed curves) at the nanometer scale. Figure 6(c) displays the line profiles of charge distribution for the 2 × 2 and √3 × √3 CDWs, taken along cuts #1 and #2 in Fig. 6(b), periods of $2a$ and $\sqrt{3}a$ are demonstrated unambiguously. Such √3 × √3 CDW puddle can be further disrupted into the 1 × 1 phase, as illustrated in Fig. S3 of SM [50]. The appearance of √3 × √3 CDW puddle and its coexistence with the 2 × 2 CDW suggests that the energies of these two phases are close; even though the ground state is 2 × 2 CDW, the √3 × √3 CDW puddle can be induced by moderate interferences. More recently, it has been found that applying high pressures on FeGe can suppress the 2 × 2 CDW gradually and eventually induce a long-ranged √3 × √3 CDW [55,56], further proving the close energies and strong phase competition between these two phases.

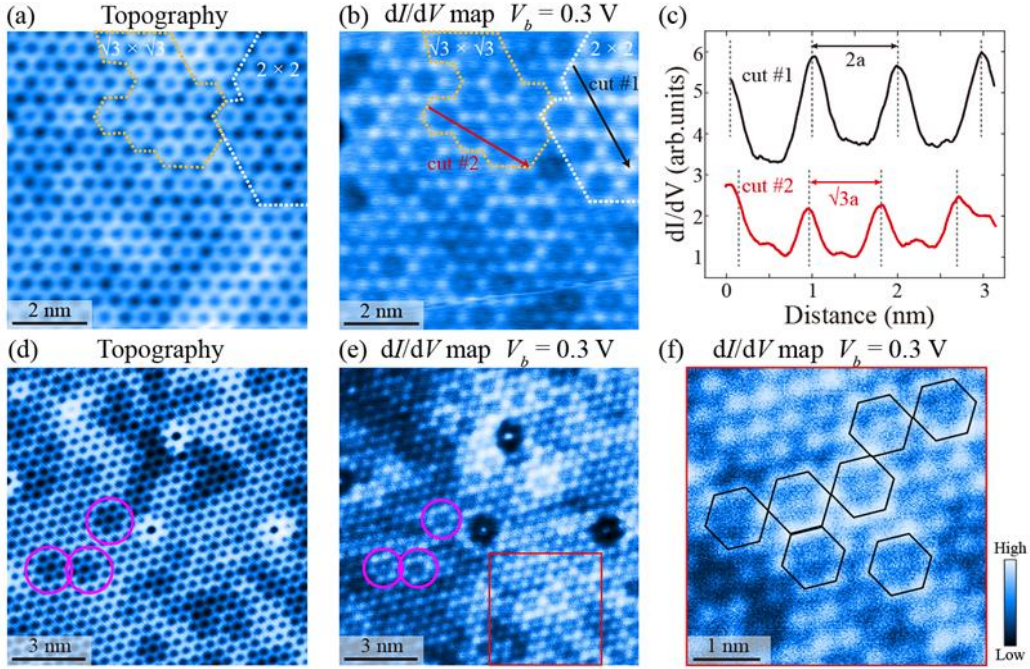

FIG. 6. Appearance of √3 × √3 charge modulations and the properties of CDW nucleus. (a),(b) Topographic image of one sample region in Fe₃Ge layer and corresponding d$I$/d$V$ map at $V_b$ = 0.3 V. Typical domains of the 2 × 2 and √3 × √3 CDWs are enclosed by the white and orange dashed curves, respectively. (c) Line profiles along cuts #1 and #2 as indicated in panel (b), exhibiting periods of $2a$ and $\sqrt{3}a$ in the 2 × 2 and √3 × √3 CDW domains, respectively. (d),(e) Detailed structure and LDOS distribution of CDW nucleus. (f) Enlarged view of the area indicated by the red box in panel (e). Magenta circles and black hexagons highlight the CDW nucleus. Measurement conditions: (a),(d) $V_b$ = 0.3 V, $I_t$ = 300 pA; (b),(e) $V_b$ = 0.3 V, $I_t$ = 300 pA, $\Delta V$ = 30 mV.

Another intriguing finding in our study is the appearance of standalone hexagonal structures in the regions where the CDW was destroyed, as highlighted by the magenta circles in Fig. 3, which look slightly lower than the surrounding lattices in topography. These hexagonal structures are far away from defects, their number increases and their positions may change during disruption processes, excluding the origin of defect-related states (Figs. S4 and S5 of SM [50]). Most of these hexagonal structures stand alone, and a few of them cluster together. Figures 6(d)-6(f) show their detailed structure and electronic properties. As highlighted by the magenta circles, an individual hexagonal structure is composed of a central Kagome cell and its six nearest neighbors, and the central cell shows a much lower LDOS than the surrounding cells (Fig. 6(e)). The LDOS pattern of an individual hexagonal structure resembles the shape of the smallest building block of the CDW (black hexagon in Fig. 6(f)), thus we conclude that each hexagonal structure is a standalone CDW nucleus. Moreover, when several such CDW nuclei aggregate together (Fig. 6(f)), they can extend along the lattice directions by sharing their corners or edges, just like those in $2 \times 2$ and $\sqrt{3} \times \sqrt{3}$ CDWs shown in Fig. 6(b). This may further hint the close energies of these two CDW phases.

To our knowledge, there are rare reports of a CDW that can be readily disrupted into standalone CDW nuclei by ordinary STM scans like $V_b \sim 1$ V, therefore, our findings suggest that the CDW in FeGe is very fragile. This phenomenon, as well as the emergence of intermediate CDW and local $\sqrt{3} \times \sqrt{3}$ CDW puddles, imply strong phase instability in FeGe due to the possibilities of multiple CDW states with close energies.

### G. Density functional theory calculations

To examine this scenario in FeGe, we performed zero-temperature density functional theory (DFT) calculations. In a recent work by one of the authors [40], a soft optical phonon mode with nearly flat dispersion around the $2 \times 2 \times 2$ CDW wave vector (red curve in Fig. 7(b)), **L**-point, has been observed in the AFM phase of FeGe at Hubbard $U = 0$ eV, which correctly produces the ordered magnetic moment ($\sim 1.5$ $\mu_B$/Fe) measured by neutron scattering experiment [37]. By increasing $U$ to induce larger spin-polarization, this phonon mode further softens by nearly the same magnitude across the **A-L-H-A** paths (Fig. 7(c)), suggesting spin-polarization may play an important role in driving this CDW. This soft phonon mode corresponds to the movements of Fe and Ge-1 atoms in the Kagome layer along $c$-axis. Therefore, the driving force of the CDW in FeGe was theoretically suggested as an enhanced spin-polarization via large dimerization of partial Ge-1 sites (labeled as Ge-1a in Figs. 7(d) and 7(e)), the subtle balance between magnetic energy saving and structural energy cost in enlarged superstructures will induce new local minimums in total energy, with very close energy to the ideal structure of FeGe [57]. The theory predicts a new global energy minimum in the $2 \times 2 \times 2$ superstructure with large dimerization of 1/4 of Ge-1 sites, which has recently been confirmed by single crystal x-ray diffraction experiments by us and others [48,51,52]. Since there are numerous ways for arrangement of the partially dimerized Ge-1a sites, i.e., numerous soft phonon modes with different wave vectors along **A-L-H-A** paths, it may induce numerous competing CDW instabilities with close energies in FeGe.

To further quantitively explain the experimental results in FeGe, we compute and compare the energies between the $2 \times 2 \times 2$ (Fig. 7(d)) and $\sqrt{3} \times \sqrt{3} \times 2$ (Fig. 7(e)) charge modulations as an example, which correspond to the soft phonon modes at **L** and **H** points, respectively. The results are shown in Fig. 7(f). Both phases are stable at small $U$ values, around $U = 0$ eV that correctly produces the ordered magnetic moments ($\sim 1.5$ $\mu_B$/Fe) measured by neutron scattering experiment for FeGe [37], the $2 \times 2 \times 2$ CDW is the ground state, but its energy is very close to the ideal Kagome structure and is only slightly lower than that of the $\sqrt{3} \times \sqrt{3} \times 2$ CDW ($\sim 0.35$ meV/atom). This can explain why the $2 \times 2$ CDW can be disrupted so easily and the appearance of $\sqrt{3} \times \sqrt{3}$ CDW in FeGe. Overall, our calculations confirm strong competition of numerous charge modulations around **L** point in FeGe, which would induce strong instability in the $2 \times 2 \times 2$ CDW ground state and result in easy CDW disruption by external interferences.

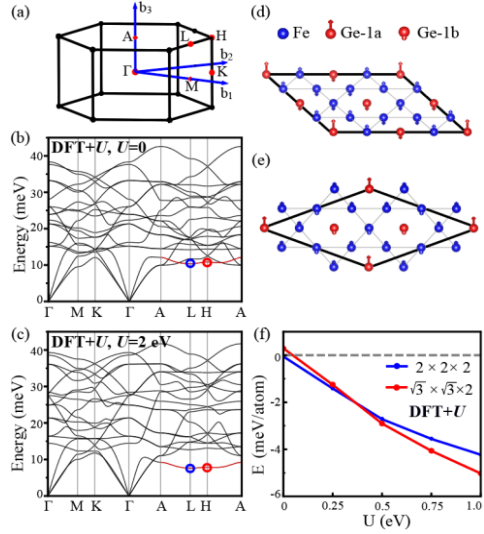

FIG. 7. A flat phonon mode softening in a large region of reciprocal space in FeGe. (a) BZ of FeGe. DFT+$U$ calculated phonon dispersion of the AFM phase of FeGe at (b) $U = 0$ eV, and (c) $U = 2$ eV. The red curve corresponds to a flat optical phonon mode that softens by nearly the same amplitude across **A-L-H-A** paths. (d),(e) In-plane $2 \times 2$ and $\sqrt{3} \times \sqrt{3}$ charge modulations, corresponding to the soft phonon mode at **L** and **H**, respectively. The arrows indicate the movements of Fe and Ge-1 atoms along $c$-axis. Longer red arrows on Ge-1a atoms indicate the partial Ge-1 sites with large dimerization (~0.65 Å), and distortions of other atoms are very small (<0.05 Å). Only a single Kagome layer is shown here, the atoms in the adjacent Kagome layer move in the opposite $c$-axis. (f) Energy of the two distorted $2 \times 2 \times 2$ and $\sqrt{3} \times \sqrt{3} \times 2$ superstructures with respect to the ideal Kagome structure (E=E$_{distorted}$-E$_{ideal-Kagome}$).

## IV. DISCUSSION

In conclusion, we find that the $2 \times 2$ CDW in FeGe, whether long-ranged or short-ranged, is very fragile and can be readily disrupted into the initial ideal Kagome structure. Small $\sqrt{3} \times \sqrt{3}$ CDW puddles are found to coexist with the $2 \times 2$ CDW in as-grown samples, and can also be induced in the intermediate process of CDW disruption, which will eventually transform into the initial $1 \times 1$ phase. Moreover, exotic intermediate CDW state and standalone CDW nuclei appear unexpectedly during the disruption process. By first-principle calculations, we find a flat optical phonon mode that softens equally in a large portion of BZ around the CDW wave vector. Furthermore, we show that the flat soft phonon mode corresponds to numerous competing CDWs with very close energies induced by large dimerization of partial Ge-1 sites, which leads to strong instability of the CDW ground state, responsible for the easy CDW disruption and the appearance of standalone CDW nucleus, $\sqrt{3} \times \sqrt{3}$ and intermediate CDWs in FeGe. Our findings provide more novel experimental aspects to understand the CDW in FeGe.

Recently, we also notice that the softening of a similar flat phonon mode and competing CDW instabilities have been observed in Kagome metal ScV$_6$Sn$_6$ [58-62], which has similar crystal structure as FeGe. The atomic distortions of the CDWs in ScV$_6$Sn$_6$ mainly consist of movements of Sc and Sn atoms along $c$-axis, while the contribution from the V Kagome layer is negligible, also similar to the case in FeGe. The band structures of FeGe and ScV$_6$Sn$_6$ changes slightly across the CDW transition, especially those associated with the Kagome layers [63-69]. This is in stark contrast to the case of CsV$_3$Sb$_5$, where the unstable phonon modes at **M** and **L** points correspond to the dominant displacement of V atoms, and reflect the intrinsic electronic instability from the V Kagome layers. These studies suggest that the intrinsic strong competing CDW instabilities may be a common feature of FeGe-type Kagome metals, and FeGe serves as a prototype to study the physics of strong competing CDW instabilities induced by structure, electronic correlations, and magnetism.

## Acknowledgments

We thank Prof. Tong Zhang for helpful discussions. Funding: This work is supported by National Natural Science Foundation of China (Grants No. 12374140, No. 12074363, No. 11790312, No. 12004056, No. 11888101, No. 11774060, No. 92065201), the Innovation Program for Quantum Science and Technology (Grant No. 2021ZD0302803), the Fundamental Research Funds for the Central Universities of China (Grant No. 2022CDJXY-002, WK9990000103), and the New Cornerstone Science Foundation. Y.-L. W. is supported by USTC Research Funds of the Double First-Class Initiative (No. YD2340002005). Author contributions: FeGe single crystals were grown by X. W. under the guidance of A. W.; STM measurements were performed by Z. C., J. Z. under the guidance of Y. Y.; DFT simulations were performed by Y. W.; The data analysis was performed by Z. C., Y. Y., Y. W., D. F., J. Z., Y. Li, R. Y., M. L., S. W.; Y. W., Y. Y. and D. F. coordinated the whole work and wrote the manuscript. All authors have discussed the results and the interpretation.